\documentclass[11pt,twoside]{article} 
\usepackage{asp2004}
\usepackage{epsf}
\usepackage{psfig}
\usepackage{lscape} 

\begin{document} 
\title{The White Dwarfs in AM CVn systems - candidates for SN\,Ia?}
\author{J.-E. Solheim}
\affil{Institute of Theoretical Astrophysics, University of Oslo,
p.o. box 1029-Blindern, N-0315 Oslo, Norway}
\author{L. R. Yungelson} 
\affil{Institute of Astronomy of the Russian Academy of Sciences, 48 Pyatnitskaya Street, 119017 Moscow, Russia}
\begin{abstract} 
 Thanks to the rapid increase of observations of Supernovae Ia,
we may now claim that the Universe is accelerating. 
 SN\,Ia are believed to be good {\it standard candles}, and 
after correcting the observations by various methods
a  precise Hubble diagram results. One of the most important problems to solve in astrophysics today is to find the progenitors of SN\,Ia. Candidates are to be found in close binary systems
 where one of the components may accumulate Chandrasekhar mass and then explode. We show that the AM CVn systems may contribute to the  SNe
population, but will not be the dominant contributor.

\end{abstract}

\section{Introduction}
One of the most surprising discoveries of the last years is that our Universe is accelerating and is not dominated by matter, but apparently
 driven apart by a dominating negative pressure, or  {\it dark energy}. This is the result of observations of distant type Ia supernovae
(SN\,Ia) which are believed to be precise distance indicators. Two independent high-$z$
 supernovae searches (Riess et al. 1998; 
 Perlmutter et al. 1999), reach the same conclusion: The SN\,Ia dim $\sim$0.29 mag at $z\sim$0.5.
 Complementary results from WMAP 
(Bennet et al. 2003) and 2dF (Peacock et al. 2001) show 
evidence for a low matter density ($\Omega_M$ = 0.3) and a non-zero cosmological
 constant ($\Omega_{\Lambda}$ = 0.7), but neither 
require the presence of dark energy (Strolger et al. 2004).\par

A SN\,Ia has no hydrogen or helium in its spectrum. The maximum
 absolute magnitude (B) varies between -18 and -20. 
It is possible to correct the light curves 
 of SN\,Ia with an  empirical formulae (parameterisation and 
stretch of the time-axes) to a final peak magnitude dispersion 
$<$ 0.2 (Fabbro 2004). The shape of the tails of the light 
curves of SNe\,Ia  are entirely explained by the nuclear
reaction
${}^{56}Ni \longrightarrow {}^{56}Co \longrightarrow {}^{56}Fe. $
However, abundance studies show that the amount of ${}^{56}Ni$
 varies from 0.07 to 0.92, which indicates that the progenitors
 must have some variety in their  composition or nuclear burning rate.
 The presence of some UV flux, the width of the peak of the early 
light curve, and the radioactive decay model, all points to a 
compact progenitor star with radius less than 10\,000 km 
(Hillebrandt \& Niemeyer 2000).\par
In the following we will discuss if AM CVn stars
 may explode as SN\,Ia  and if  they can contribute significantly to the SNe population.

\section{The progenitor requirements}

The progenitors of SNe Ia are not known for certain. 
One of the most widely accepted models is that of 
a carbon/oxygen (CO) white  dwarf accreting mass until it reaches
$\sim$1.378$M_{\sun}$ (Nomoto, Thielemann \& Yokoi 1984), close to the 
Chandrasekhar limit, and then explodes as a SN\,Ia.
In more general terms a SN\,Ia model 
must fill the following requirements (Hillebrandt \& Niemeyer 2000):

$\bullet$ Agreement of ejecta composition and velocity with the observed spectra and light curves;

$\bullet$ Robustness of the explosion mechanism (no fine tuning needed);

$\bullet$ Intrinsic variability accepted --- at least one parameter to explain the variations in maximum energy output;

$\bullet$ Correlation with progenitor systems and their evolution.

The {\sl Hubble} Higher $z$ Supernova Search (Strolger et al. 2004) 
has discovered 42 SNe in the redshift range $0.2 < z <1.6$ and compared the number counts of SNe in
$\Delta z$ bins with the expected distributions based on the 
following model parameters: 
 {\it delay time} and progenitor distribution or birth rate. 
Strolger et al. (2004) get the best fit to observations for a narrow Gaussian ($\Delta \tau = 0.2\tau$) starbirth model with
mean delay times $\tau \approx$  3--4 Gyrs.  \par
The challenge is now to find progenitor populations that match the observed redshift distribution.

\begin{figure}[!ht]
\plotfiddle{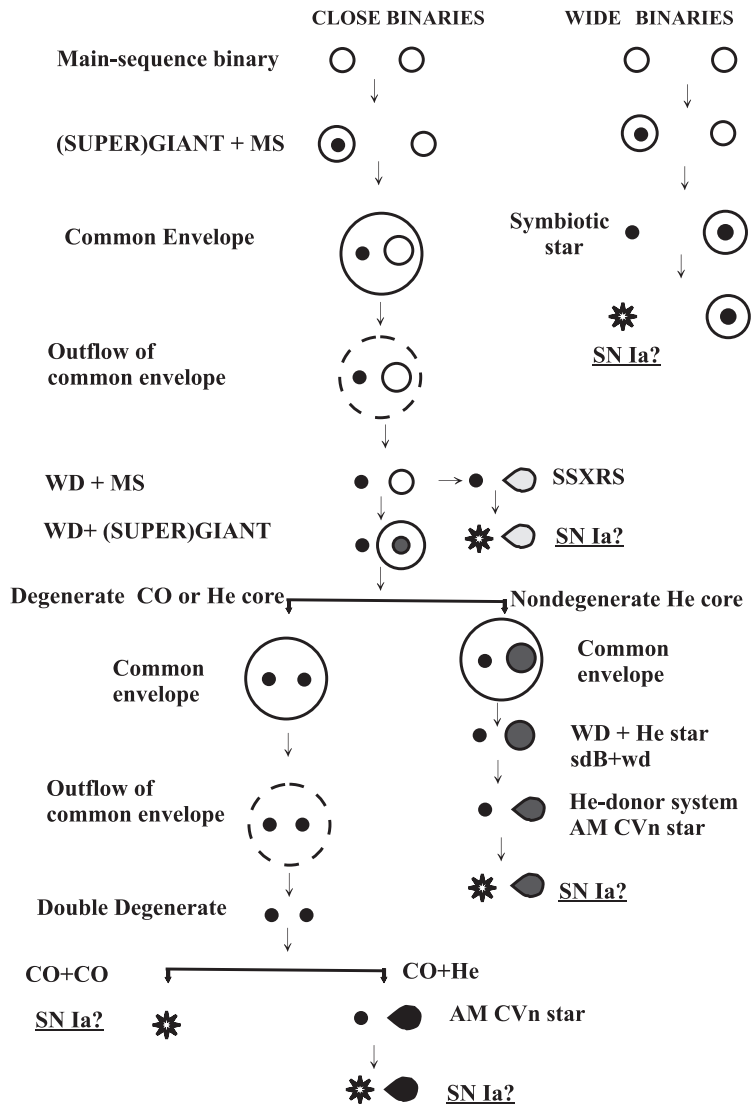}{15cm}{0}{55}{55}{-180}{-15}
\caption{Evolutionary scenarios for potential progenitors of SN\,Ia. The lower left branch shows the double degenerate (DD) option where a pair of CO white dwarfs merges. The lower right branch (CO+He) represents the double-degenerate family of AM~CVn stars. 
The Helium-star family of AM~CVn stars is formed in the branch 
of objects with nondegenerate helium cores.
}
\end{figure}

\section {Progenitors for SN\,Ia}

\begin{figure}[!ht]
\plotfiddle{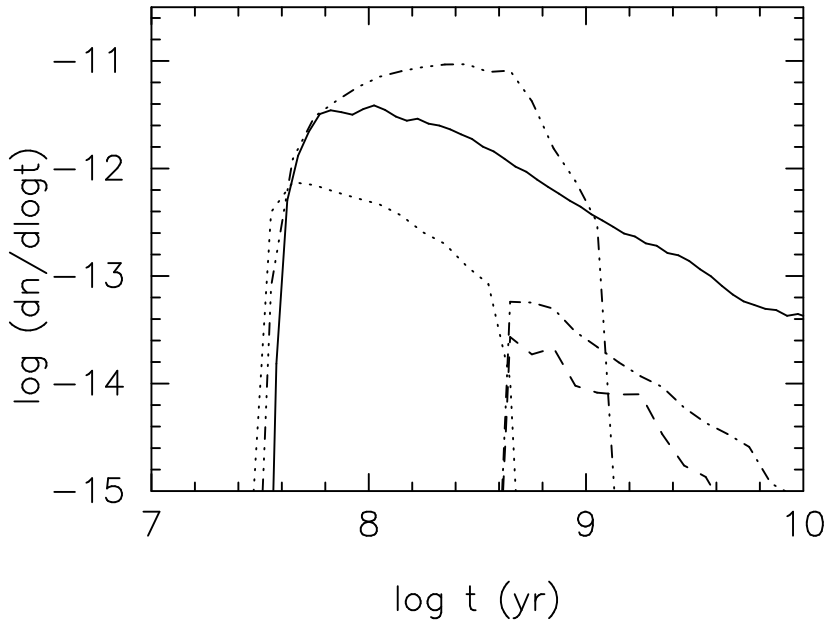}{8cm}{270}{50}{50}{-210}{280}
\caption{
Rates of potential SNe\,Ia after an instantaneous star
formation burst.
Solid line --- merger of CO+CO pairs with total
mass $ > M_{Ch}$.
Long dashes --- the  rate of accumulation of $M_{Ch}$ in AM CVn systems
of DD-family when mass loss during helium
flashes is taken into account.
Dots ---  the same in binaries with He-star donors.
Dash-dot-dot-dot line ---  ELDs in binaries with
He-star donors.
Dash-dots --- accumulation of $M_{Ch}$ in AM CVn systems
of DD-family  if 100\%
efficient accumulation of accreted He is assumed.
The scaling corresponds to formation of one binary with mass
of the primary component $> 0.8 M_{\sun}$ per
year, a flat distribution over initial
separation of components, and a flat distribution of the
mass ratios of the components.
}
\end{figure}

We can visualize in Figure 1, four main close binary channels 
for creating SN\,Ia:

$\bullet$ Mergers of double degenerates resulting in the formation of 
a $M \ga M_{Ch}$ object and central carbon ignition (the CO+CO  DD channel);

$\bullet$ Accumulation of Chandrasekhar mass by CO dwarf via stable Roche lobe overflow in semidetached systems with He-white dwarf; 

$\bullet$ Accretion of helium from a non-degenerate He-rich companion at a rate of $\dot M \sim 10^{-8}M_{\sun}$yr$^{-1}$, resulting in the accumulation of a He layer of $\sim(0.10-0.15)M_{\sun}$. The ignition of the carbon in the core is induced by a detonation in the He-layer
(Edge Lit Detonation, ELD). If for some reason, e. g., lifting effect of rotation, helium ignition is mild and does not result in disruption of the donor (helium Nova?), 
the surviving system may become a He-family AM~CVn star with a
semi-degenerate secondary; 

$\bullet$ Mass transfer from a H-rich main-sequence or subgiant star, forming a supersoft X-ray source (SSXRS) which may 
produce SN\,Ia via ELD or accumulation of $M_{Ch}$. 


 In Figure 1, we also show a wide binary channel, with a white dwarf
 that may accumulate $M_{Ch}$ in a symbiotic  system via wind accretion. However, it is estimated that it gives only a minor contribution to the total rate of SN\,Ia.

\section {Can AM~CVn objects contribute to the SN I population?}

\begin{figure}[!ht]
\plotone{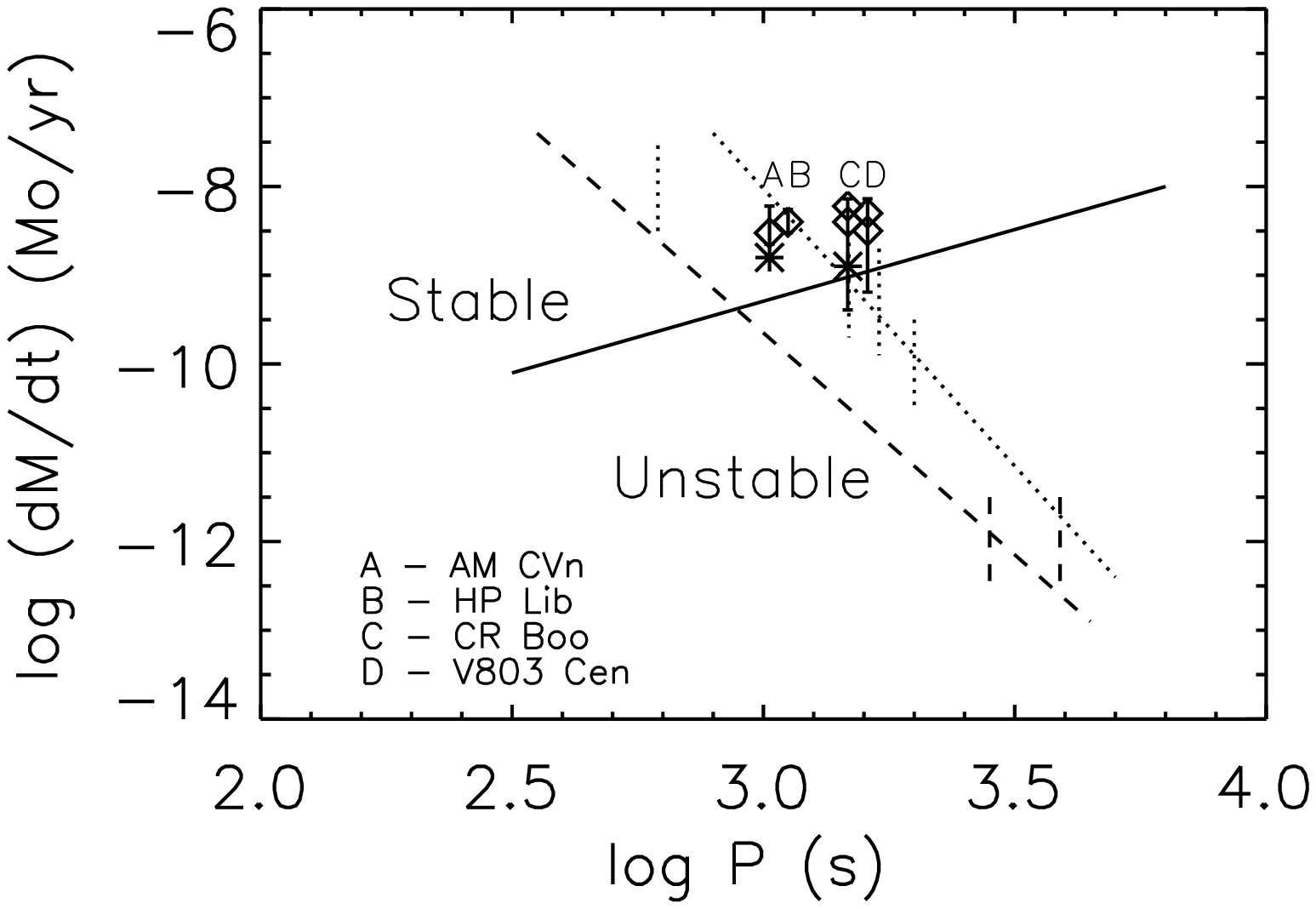}
\caption{The mass transfer rate $\dot M$ versus period diagram for 
AM~CVn objects. The evolution of the DD and helium star family of  AM~CVns are shown as broken or dotted lines. For four objects the mass
transfer rates are determined from modeling of disk spectra (stars and diamonds with bars -- where the bars indicate 
the variation in mass transfer from low to high state).  For the rest of the objects $\dot M$s are suggested by vertical bars. The solid line from the lower left to the upper right, defines the border
between stable and unstable disks (Tsugawa \& Osaki 1997)}
\end{figure}

In Figure 1 we find AM~CVn stars in two channels. The main characteristics of AM~CVn systems are that they are 
ultra-compact binaries, with orbital periods 10-65 min, showing no hydrogen in their spectra. A dozen such systems are known (Solheim 2003; Nelemans 2004).
They may have degenerate (DD-family) or semi-degenerate  
(He-star family) donors. In both families their progenitors
evolve with the orbital period shortened by loss of angular momentum
 by
gravitational wave radiation, to a certain minimum period.
In the course of further evolution the orbital period increases,
while the mass transfer rate is decreasing
(see, e. g., Nelemans et al. (2001)).
In both channels the accretor may grow as mass is transferred, and when reaching the Chandrasekhar limit explode or collapse. In the case of an explosion, it will be a SN\,Ia. In the case of collapse,  
the formation of a neutron star is expected, but observational manifestation of such an event is unclear.
In the DD-family, the population mainly depends on the efficiency of
 the tidal coupling between the accretors spin and orbital motion.
For the He-star family the population depends on the chances for
an ELD to occur before the Chandrasekhar mass is reached (Nelemans et al. 2001).\par

Using an updated version of the Tutukov \& Yungelson (1996) code for population of stars with helium donors, we have computed the 
occurrence rate of SN\,Ia  appearing from  various channels. 
Figure~2 shows
the rates after an instantaneous star 
formation burst.
For AM CVn stars we apply two options: efficiency of He accumulation 
as determined by 
Iben \& Tutukov (1996) and a 100\% efficiency that provides us the upper limit for occurrence rate.
We find that even in the most optimistic case for production of SNe from AM CVns, the CO+CO channel dominates,  but it does not follow the distribution  observed in the {\sl Hubble} Higher $z$ Supernova Search 
(Strolger et al. 2004). 

In Figure 3 we present the population of known AM~CVn systems in a $\dot M$ vs. period diagram, where also the evolutionary tracks for 
the objects of  DD-  and He-star families are shown.
 For  only 4 objects we have reliable $\dot M$ determinations (Nasser et al. 2001; El-Khoury \& Wickramasinghe 2000), the others 
have uncertain values based on periods, mass ratios and luminosties.  
The figure suggests that most of the currently observed
 AM~CVns are likely to belong to the He-star family, and this may mean that progenitors of this population are not decimated by ELDs, in agreement with conclusions by Yoon \& Langer (2004),
who find that the effects of rotation reduce the violence of helium flashes. We may speculate that helium Novae are produced instead of ELDs. On the other hand, we may also speculate about a sink for the DD-family, either by inefficient coupling between spin and 
orbital motion (Nelemans et al. 2001), or by mergers
in common envelopes associated with shell He explosions at relatively high $\dot M$ or expansion when the accretion rate is
 $\dot M_{Edd} >  \dot M_{acc} > \dot M_{crit}$,  where $M_{crit}$ is the upper limit for the rate of stable He burning.

\section {Conclusion}

The AM~CVn systems we know today are helium transferring objects,
 which can explode as SN\,Ia. 
However, with accreator masses of the order 1.0 $M_{\sun}$, 
donor masses $< 0.1 M_{\sun}$ (Solheim 2003) and a low mass transfer rate, 
the majority of currently observed AM\,CVns
 will not reach the $M_{Ch}$ limit in a Hubble time.
\par
The apparent lack of AM CVns of the DD-type, suggests that
either the efficiency of formation of AM CVns in this family 
is lower than current calculations suggest, or a fraction
 of this population has not been  
 observed because of yet unrecognized selection effects.
In any case {\it at present the AM~CVn stars can contribute at most 
 $\sim  1$ per cent} to the SN\,Ia occurrence rate, and 
 they do not follow the number density pattern determined 
by the {\sl Hubble} Higher $z$ Supernova Search.\par
However, the study of AM~CVns may give us valuable information about
 their progenitors and conditions for SN\,Ia explosions, which are needed to explain the existence of 
standard candles and the need for a dark energy in our Universe. 

\acknowledgements{LRY is supported by RFBR grant no. 03-02-16254 and Federal Program ``Astronomy''.}


\begin{references}
\reference Bennet, C. L., Halpern, M., Hindshaw, G. et al., 2003, \apjs, 148, 1
\reference El-Khoury, W., \& Wickramasinghe, D., 2000, \aap, 358, 154
\reference Fabbro, S., 2004, \apss, 290, 1 
\reference Hillebrandt, W. \& Niemeyer J. C., 2000, \araa, 38, 191
\reference Iben, I. Jr. \& Tutukov, A.V., 1996, \apjs, 105, 145 
\reference Nasser, M. R., Solheim, J.-E., \& Semionoff, D., 2001, \aap, 373, 222
\reference Nelemans, G., Portegies Zwart, S. F., Verbunt, F., \& Yungelson, L. R., 2001, \aap, 368, 939
\reference Nelemans, G., 2004, astro-ph/0409676
\reference Nomoto, K., Thielemann, F., \& Yokoi, K., 1984, \apj, 286, 644
\reference Peacock, J. A., Cole S., Norberg, P. et al., 2001, \nat, 410, 169
\reference Perlmutter, S., Aldering, G., Goldhaber, G. et al. 1999, \apj, 517, 565
\reference Riess, A. G., Filippenko, A. V., Chalis, P. et al., 1998, \aj, 116, 1009
\reference Solheim, J.-E., 2003 in {\it White Dwarfs} eds D. Martino et al., NATO Sci. Ser. II, Kluwer, v. 105, p 299
\reference Strolger, L.-G., Riess, A. G., Dahlen T. et al., 2004, \apj, 613, 200
\reference Tsugawa, M. \& Osaki, Y., 1997, \pasj, 49, 75
\reference Tutukov, A. V. \& Yungelson, L.R., 1996, \mnras, 280, 1035
\reference Yoon, S.-C. \& Langer, N., 2004, \aap,  419, 645
\end{references}
\end{document}